\documentstyle[12pt,epsfig]{article}
\def\lsim{\mathrel{\raise3pt\hbox to 8pt{\raise -6pt\hbox{$\sim$}\hss{$<$}}}}
\newcommand{\bpi}{\mbox{\boldmath $\pi$}}
\newcommand{\btau}{\mbox{\boldmath $\tau$}}

\def\bt{{\bf t}}
\def\haf{\textstyle{1\over2}}
\def\minus{\mbox{$-$}}
\newcommand{\vr}{\vec{r}}
\newcommand{\vk}{\vec{k}}
\newcommand{\vx}{\vec{x}}
\newcommand{\vq}{\vec{\, q}}
\newcommand{\vl}{\vec{l}}
\newcommand{\vsig}{\vec{\sigma}}
\newcommand{\vnabla}{\vec{\nabla}}
\newcommand{\fpi}{f_{\pi}}
\newcommand{\mpi}{m_{\pi}}

\newskip\humongous \humongous=0pt plus 1000pt minus 1000pt
\def\caja{\mathsurround=0pt}

\newif\ifdtup
\def\panorama{\global\dtuptrue \openup1\jot \caja
        \everycr{\noalign{\ifdtup \global\dtupfalse
        \vskip-\lineskiplimit \vskip\normallineskiplimit
        \else \penalty\interdisplaylinepenalty \fi}}}
\def\eqalignno#1{\panorama \tabskip=\humongous
        \halign to\displaywidth{\hfil$\displaystyle{##}$
        \tabskip=0pt&$\displaystyle{{}##}$\hfil
        \tabskip=\humongous&\llap{$##$}\tabskip=0pt
        \crcr#1\crcr}}
\begin{document}

\begin{center}

{\Large {\bf Charge-Symmetry Breaking and the Two-Pion-Exchange
Two-Nucleon Interaction}}\\

\vspace*{0.4in}
{\bf J.\ L.\ Friar} \\
{\it Theoretical Division,
Los Alamos National Laboratory \\
Los Alamos, NM  87545} \\
\vspace*{0.10in}
\vspace*{0.10in}
{\bf U.\ van Kolck}\\
{\it Department of Physics,
University of Arizona\\
Tucson, AZ 85721} \\
and \\
{\it RIKEN-BNL Research Center,
Brookhaven National Laboratory\\
Upton, NY 11973}\\
\vspace*{0.10in}
\vspace*{0.10in}
{\bf G.\ L.\ Payne}\\
{\it Department of Physics and Astronomy,
University of Iowa\\
Iowa City, IA 52242}\\
\vspace*{0.10in}
\vspace*{0.10in}
{\bf S.\ A.\ Coon}\\
{\it Department of Physics,
New Mexico State University\\
Las Cruces, NM  88003}\\
and\\
{\it Division of Nuclear Physics,
U.S. Department of Energy\\
SC-23, Germantown Building,
1000 Independence Avenue\\
Washington, D.C. 20585-1290}\\
\end{center}
\vspace*{0.10in}
\begin{abstract}
Charge-symmetry breaking in the nucleon-nucleon force is investigated
within an effective field theory,
using a classification of isospin-violating interactions 
based on power-counting arguments. 
The relevant charge-symmetry-breaking interactions corresponding to the 
first two orders in the power counting are discussed,
including their effects on
the $^3$He $\minus$ $^3$H binding-energy difference.
The static 
charge-symmetry-breaking potential linear in the nucleon-mass difference is 
constructed using chiral perturbation theory. 
Explicit formulae in momentum and configuration spaces are presented. 
The present work completes previously obtained results. 
\end{abstract}

\pagebreak

\section{Introduction}

Significant advances in understanding isospin violation in the nuclear
force have been made in the past decade. Experimental progress (which is
reviewed and summarized in Refs.\cite{iv1,iv2,iv3}) has been supplemented
recently by the advent of chiral perturbation theory
(ChPT)\cite{weinberg,iv,texas,2NTPE, BvK}. 
This powerful technique casts the symmetries
of QCD into effective interactions of the traditional, low-energy degrees of
freedom of nuclear physics (viz., nucleons and pions). 
These building blocks (in
the Lagrangian) can then be combined in a systematic way to produce nuclear
forces that violate isospin in exactly the same way as in QCD.

An important feature of effective field theories is power counting, 
which is the technique used to organize
calculations\cite{weinberg,iv,texas,BvK,ndpc,dpc}. 
A well-defined ordering of
terms in the Lagrangian according to scales intrinsic to QCD and nuclei is used
to generate all terms of a particular size.  
Terms in the Lagrangian are labeled
(in the conventional way, ${\cal L}^{(n)}$) 
by the number of implicit powers ($n$) of the inverse
of the large-mass scale of QCD, $\Lambda \sim$ 1 GeV.
When considering processes where the typical momentum is of the order
of the pion mass, $Q \sim m_\pi$, amplitudes are
expanded in powers of $Q/\Lambda$.

Much work has been done in the past few years regarding
the derivation of the isospin-symmetric part of the 
nuclear potential\cite{BvK}. 
The components of short ($\sim \Lambda^{-1}$) range in the potential
come from (renormalized) contact interactions
whose strengths are not determined by symmetry.
The components of longer range arise from pion exchanges,
and are determined in terms of pion couplings to the nucleon,
which are fewer and in many cases determined from
processes involving a single nucleon.
Perhaps the most significant result of this method is
the first derivation of a two-pion-range potential
consistent with the approximate chiral symmetry of 
QCD\cite{iv,texas,2NTPE}.
This two-pion-exchange potential (TPEP) has been incorporated
in the Nijmegen phase-shift analysis\cite{nijmegen,rob_mart}.
Pion-nucleon couplings determined from
two-nucleon data were found to be in
good agreement with other determinations
based on pion-nucleon scattering.

In effective theories, isospin-breaking interactions 
can be classified\cite{iv} according to whether their
origin is the mass difference between $u$ and $d$ quarks 
or hard electromagnetic (EM)
interactions at the quark level. The soft
EM interactions (such as the
Coulomb force) 
can be constructed in the usual way\cite{av18,pig}.

The power counting for isospin-violating interactions was developed by
one of us\cite{iv}, and it explains the sizes
of the various isospin structures present in the nuclear force.  
A convenient
and universal\cite{iv1} classification for nuclear isospin is:  class (I) -
isospin conserving; class (II) - charge-independence breaking (CIB) of isospin,
but charge symmetric; class (III) - charge-symmetry breaking (CSB) of isospin;
class (IV) - isospin mixing in the $np$ system between $T=0$ and $T=1$. 
Power
counting can be used to demonstrate\cite{iv} that 
a class (N) force appears only at order $n=N-1$ or higher.
We therefore deduce on the basis of QCD a result that was noticed
before on an empirical basis:
class (I) forces are stronger
than class (II), which are stronger than class (III), which are stronger than 
class (IV).  
 
Using this formalism, several of the various isospin components 
of the nuclear potential have been calculated\cite{iv,1loop,pig,FvK,csb1,csb2}.
In particular, we have computed the long-range components of the
class (I)\cite{weinberg,iv,texas,2NTPE}
and class (II)\cite{iv,1loop,pig,FvK} 
two-nucleon potentials, up to order $n=3$.
The class (III) isospin violation is the purview of this work.

Charge symmetry\cite{cs-rev} can be loosely defined as invariance under turning
neutrons into protons and protons into neutrons. 
It has long been considered a particularly interesting
aspect of isospin violation because the difference
in quark masses is a source of CSB.
On the other hand,
a significant interaction
that violates this symmetry is the electromagnetic interaction, which is large
between two protons and very small between two neutrons. Moreover, the Coulomb
interaction between protons is long-ranged, while it has a nuclear range for
two neutrons and is therefore
indistinguishable from nuclear interactions. There are corrections
of order $1/M^2$ to the static Coulomb interaction (where $M$ 
is the nucleon mass), which are part of the Breit interaction. 
Although these familiar
electromagnetic interactions dominate the CSB in nuclei, we will largely ignore
them and concentrate on the nuclear mechanisms\cite{1loop}. 
In the nuclear force
the comparison that interests us is the difference in forces 
between two protons
and two neutrons, which is restricted to the $T=1$ channel for two nucleons.

The longstanding interest in CSB\cite{cs-rev} has been 
highlighted recently by two exciting new experiments.
One experiment, carried out at TRIUMF and currently undergoing final stages
of analysis, measures the front-back asymmetry of the pion produced
in the reaction $np\to d\pi^0$ close to threshold\cite{Allena}.
Another experiment has just been completed at IUCF,
and measures the near-threshold cross section
for the reaction $dd\to \alpha\pi^0$\cite{EdAndy}.
It has been argued in Ref.\cite{vKMN} that the chiral properties of
two different contributions to the nucleon-mass difference
can give a relatively large effect in $np\to d\pi^0$,
of opposite sign to more well-known mechanisms.
A similar phenomenon might exist
in $dd\to \alpha\pi^0$\cite{vKMN}.
There is reason to hope that these data will allow 
a model-independent (albeit crude) determination of 
both quark-mass and electromagnetic components of the
nucleon-mass difference.

An issue that arises naturally is the role of 
the nucleon-mass difference in the two-pion-exchange
nucleon-nucleon potential. 
Ref.\cite{csb1} calculated the crossed-box contribution, and 
it contains a comprehensive discussion of the 
older literature\cite{olderlit,pi-eta},
where aspects of chiral symmetry are not emphasized. 
More recently, Ref.\cite{csb2} calculated a CSB seagull contribution
using the formalism of Ref.\cite{iv}.

In this paper we discuss the relative sizes of
CSB interactions in effective field theories,
and calculate the two-nucleon potential of two-pion
range that arises 
from the nucleon-mass difference.
Pieces of this potential have appeared before\cite{csb1,csb2}.
Here we complete the calculation of the long-range component
of the class (III) potential
up to (and including) order $n=3$.

\section{Sizes}

In the chiral Lagrangian,
the largest isospin-violating terms are those that
determine the mass differences of nucleons and pions\cite{iv}. 
In order to compare these and various other contributions,
it is useful to estimate relative sizes of EM and quark-mass effects. 
The pion-mass splitting provides a good starting point.

The pion-mass splitting
(squared), $\delta m_{\pi}^2 = (m_{\pi}^{\pm})^2 \minus (m_{\pi}^{0})^2$, 
receives contributions from both 
the difference between the masses of the $u$ and $d$ quarks
and 
EM interactions.
Because of the chiral transformation properties of the pion,
the contribution from the quark-mass difference
is proportional to $(m_{\pi}^2/\Lambda)^2$.
The proportionality constant ($\epsilon^2$) is on the order of the
square of the difference of quark masses divided by their sum 
($\epsilon \sim 0.3$).
The pion-mass difference is therefore mostly due to EM interactions.
It is nominally of order $n=\minus 2$, 
but the large strength implied by this  counting is
compensated by the small fine-structure constant ($\alpha \sim 1/137$),
which reduces the strength by
slightly more than two orders of magnitude. 
The size of the pion-mass splitting
is then rather accurately described by 
$\delta m_{\pi}^2 \simeq \alpha \Lambda^2 / \pi$, 
with
$\Lambda \sim m_{\rho}$, the mass of the $\rho$-meson\cite{iv}. This suggests
that the relevant dimensionless parameter for the EM-induced isospin violation
is $\delta m_{\pi}^2/\Lambda^2 = \alpha/\pi \cong 2 \cdot 10^{-3}$, which
numerically is close to 
$\epsilon m^3_{\pi}/\Lambda^3 \sim 2 \cdot 10^{-3}$.
Since the latter would ordinarily
correspond in power counting to $n=3$,
we deduce a convenient
mnemonic of adding 3 to the order of the EM-induced isospin-violating 
Lagrangian
when comparing sizes with quark-mass-induced mechanisms. 
Henceforth our power counting for EM-induced interactions will contain this 
additional factor of 3 (e.g., $n=-2$ as counted above will be counted as $n=1$ 
below).

The power counting has been used in Ref.\cite{iv} to classify
various contributions to the nuclear potential.
The leading isospin-breaking interaction appears at $n=1$, when the
pion-mass difference is inserted in the one-pion-exchange potential (OPEP).
It gives a class (II) force, 
which is of relative strength $\delta m_{\pi}^2/m_{\pi}^2$
(times OPEP)\cite{iv}. 
Many other class (II) forces appear up to order $n=3$.
There are three EM mechanisms of roughly the same relative strength,
$\alpha/\pi$ (times the usual OPEP) 
produced by $n=3$ terms in the Lagrangian or from loops.
They are:
the two-pion-exchange
potential with different charged and neutral pion masses\cite{FvK}, 
CIB in the pion-nucleon coupling constant\cite{1loop} and consequently in OPEP,
and the $\pi-\gamma$-exchange potential\cite{pig}.
Moreover, there are OPEP corrections 
that arise from higher-order isospin-conserving pion-nucleon interactions
(such as recoil in the usual pion-nucleon coupling),
and from
second-order effects due to the 
the pion- and nucleon-mass differences\cite{iv}.

Here we concentrate on the class (III) mechanisms.
It is worthwhile 
to discuss briefly the various mechanisms that
contribute to CSB in the $NN$ system and in the trinucleon system 
($^3$He $\minus$
$^3$H), their relative sizes, and their relationship in the context of power
counting. 

The bulk of the CSB is due to the soft EM interactions. The 764 keV
binding-energy difference of $^3$He and $^3$H is largely accounted for by the
648 keV Coulomb energy difference\cite{coulomb} plus approximately 29 keV from
the small Breit corrections plus 
vacuum polarization\cite{brandenburg,av18, sasakawa,
argonne, nogga}. Calculations of these contributions appear to be rather 
robust,
and we henceforth ignore them. The remaining mechanisms that generate
approximately 87 keV are contained in ChPT, and we discuss them below.

The most obvious signature of CSB due to nuclear forces is the difference in 
the
scattering lengths of two neutrons and two protons, once all EM mechanisms are
accounted for and the effect of the different nucleon masses on the kinetic
energy (see Eqn.~(1c) below) is treated. The resulting ``experimental''
$a_{nn}\minus a_{pp}$ scattering-length difference\cite{iv2,iv3,1loop} of
$-$1.5(5) fm is then attributed to CSB in the nuclear force. 
Typically one treats
only the short-range (as opposed to pion-range) CSB force, by adjusting the
force to produce the desired difference in scattering lengths. This CSB
modification of the force then produces a contribution to the $^3$He $\minus$
$^3$H binding-energy difference of approximately 65(22) keV, a number that also
appears robust\cite{argonne,nogga,csmodel} 
and can accommodate the 87 keV
binding-energy difference missing after soft EM processes are taken into
account. Note that this scattering-length difference is generated by the total
CSB nuclear force, and does not differentiate between components of different
ranges.

The mass of a single nucleon can be expressed
in terms of isospin operators
(the Pauli isospin operator $\btau$ satisfies
$\btau = 2\, \bt$, where $t_z$ gives +1/2 for a proton 
and $-$1/2 for a neutron):
$$
M_N = \haf (M_p + M_n) + \haf (M_p - M_n) \tau_z \equiv M +\haf \delta M_N 
\tau_z\, , \eqno(1a)
$$
where $M = \haf (M_p + M_n)$ is the average nucleon mass, $\delta M_N = (M_p -
M_n)$ is the nucleon-mass difference, and $\haf \delta M_N \tau_z$ is a CSB
nucleon-mass interaction.
The nucleon-mass
difference, $\delta M_N$, receives contributions
from both the quark-mass difference, $\delta M_N^{\rm qm}$,
and EM interactions at the quark level, $\delta M_N^{\rm EM}$,
$$
\delta M_N = \delta M_N^{\rm qm} + \delta M_N^{\rm EM} \, . \eqno (1b)
$$
(We neglect here mixed contributions of higher order.)
The piece due to the quark-mass difference 
is the sole CSB contribution of order $n$=1, implying that this
quantity is proportional to $m_{\pi}^2/\Lambda$.
The parameter $\epsilon m^2_{\pi}/\Lambda^2 \sim  10^{-2}$ plays 
an important
role in the power counting for isospin violation, and would lead to a
nucleon-mass difference on the order of 8 MeV (a factor of three 
higher than most model estimates, and therefore reasonable).
The corresponding nucleon-mass term from EM interactions at the quark level is
of order $n = 2$, but is numerically comparable 
to the quark-mass term (somewhat smaller and of
opposite sign) because it implicitly contains a power of the fine-structure
constant. 

The largest CSB interactions in the Lagrangian thus come
from the nucleon-mass difference of orders 
$n$= [1,2]
for the two different mechanisms. 
The effects of the nucleon-mass difference
on nuclear amplitudes are somewhat suppressed, however.
Summing Eqn.~(1a) over all nucleons produces an overall 
constant
(determined by the average nucleon mass), which can be removed by a shift in 
the
zero of energy, and a term proportional to the $z$-component of the total
nuclear isospin, $T_z$. The latter term will contribute to a CSB shift in the
nuclear kinetic energy (for simplicity we restrict ourselves to 
only
two nucleons in their center-of-mass (CM) frame)
$$
\delta T = - T \frac{\delta M_N}{2 M} T_z\, , \eqno(1c)
$$
where $T$ is the usual CM kinetic energy for two nucleons with equal masses, 
and
$T_z = t_z (1) + t_z (2)$ for two nucleons. The CSB nucleon-mass difference 
will
also lead to a modification of the nuclear potential energy. The CSB part of
Eqn.~(1a) ($\delta M_N T_z$) will commute with an isospin-conserving 
Hamiltonian
and therefore will play no role to order $\delta M_N$, except inside certain
loops (viz., crossed-box and triangle diagrams) that involve exchanging two
charged pions. In the latter case for a process initiated by two identical
nucleons of one type, the intermediate state would contain two nucleons of the
${\bf other}$ type. Exchanging neutral mesons clearly does not invoke this
mechanism, nor does ladder approximation (sequential exchanges of neutral
mesons) for two protons or two neutrons. Because this effect requires a loop, 
it
is of order $n=3$. Thus although nominally of order $n=1$ (by itself), the
nucleon-mass difference actually first contributes in the much higher order
$n=3$, which is a reduction in size typical of loop contributions. In addition
to these loop insertions, there are associated triangle-graph interactions from
seagulls (discussed below).

Other contributions in the Lagrangian begin at order $n=2$ for the quark-mass
terms (i.e., they are the largest) and arise from two types of CSB nuclear
forces: short-range forces (such as those arising from 
$\rho - \omega$\cite{rho-w} and $a_1 - f_1$\cite{a1-f1} mixing), 
and from isospin violation in the pion-nucleon coupling
constant (a model for the latter is provided 
by $\pi - \eta$ mixing\cite{pi-eta}). 
We denote the former by $V^{\rm CSB}_{\rm SR}$ and the latter by
$V^{\rm CSB}_{\pi}$. To leading order\cite{iv} for two nucleons one has
$$
V^{\rm CSB}_{\rm SR} = (\gamma_s +\bar\gamma_s)\, 
V_0 (r)\,T_z + (\gamma_{\sigma}+\bar\gamma_{\sigma})\, V_1 (r)\; 
\vsig(1) \cdot \vsig(2)\, T_z \, , \eqno(2)
$$
where $\vr$ is the separation of nucleons 1 and 2, 
and $V_0$ and $V_1$ have unit
volume integral. The constants $\gamma_s$ and $\gamma_{\sigma}$ 
stem from the quark-mass difference and are of order
$\epsilon m^2_{\pi}/f_{\pi}^2 \Lambda^2$,
while $\bar\gamma_s$ and $\bar\gamma_{\sigma}$
are EM corrections of order $\alpha/\pi f_\pi^2$.
This implies a potential strength
$V^{\rm CSB}_{\rm SR} \sim (\epsilon m_{\pi}^2/\Lambda^2) V_{\rm SR}^{NN}$,
where $V_{\rm SR}^{NN}$ is the expectation value of the short-range interaction
between either the $pp$ pair in $^3$He or the $nn$ pair in $^3$H. A 34-channel
calculation using the AV18 potential\cite{av18} gives 
$-$7.6 MeV for the latter,
or an estimate of about 45 keV for the contribution of the CSB short-range
interaction to the $^3$He $\minus$ $^3$H binding-energy difference. This
interaction has traditionally always been a part of CSB studies.

The OPEP contains a CSB (coupling constant) modification\cite{iv,1loop} 
$$
V_{\pi} = v_{\pi}\; ( 
\bt (1) \cdot \bt (2) -\frac{(\beta_1+\bar{\beta}_3)}
{2 g_A}\, T_z ) \equiv V_{\pi}^0 + V_{\pi}^{\rm CSB} \, , \eqno(3)
$$
where $V_{\pi}^0 = v_{\pi} \, \bt (1) \cdot \bt (2)$ is the usual
isospin-conserving OPEP, and we do not write (higher-order) recoil corrections
explicitly.
Here $\beta_1 \sim \epsilon m^2_{\pi}/\Lambda^2$ 
is the (quark-mass induced) CSB pion-nucleon coupling
constant and 
$\bar{\beta}_3 \sim \alpha/\pi$
is the (EM-induced) CSB pion-nucleon coupling constant. 
The size of $V_{\pi}^{\rm CSB}$ is constrained only by
an upper limit on $\beta_1 + \bar{\beta}_3$,
arising from the Nijmegen phase-shift analysis of $NN$ data\cite{1loop}.
A numerical estimate 
of $V^{\rm CSB}_{\pi}$ in the trinucleon system can be obtained from Eqn.~(3), 
which shows
that the contribution to the trinucleon binding-energy difference is
approximately $\langle V^{\rm CSB}_{\pi}\rangle \cong -4 (\beta_1 +
\bar{\beta}_3)\, V^{NN}_{\pi}/g_A$, where $V^{NN}_{\pi}$ is the expectation
value of the potential energy due to OPEP between either the 
$pp$ pair in $^3$He
or the $nn$ pair in $^3$H. A 34-channel Faddeev calculation using the AV18
potential\cite{av18} produces $-$1.67 MeV for the latter. 
Using the experimental
value of $(\beta_1 + \bar{\beta}_3) = 0(9)\cdot 10^{-3}$ from 
Ref.\cite{1loop}
produces $\langle V_{\pi}^{\rm CSB}\rangle \sim$ 0(50) keV of uncertain sign.
The dimensional estimate\cite{iv,1loop} using 
$\beta_1 \sim \epsilon m_{\pi}^2/\Lambda^2 \sim 10^{-2}$ 
produces $\langle V_{\pi}^{\rm CSB}\rangle \sim$ 50 keV, 
and also of uncertain sign. 
Thus a substantial part
of the ``short-range'' contribution to CSB (65(22) keV) in the
trinucleons could come from CSB in OPEP, rather than from a shorter-range
interaction. In any event the two mechanisms are predicted by power counting to
be the dominant hadronic contributions and to be roughly comparable in size.

The next order in ChPT for CSB is $n=3$. The modification of 
the nuclear kinetic
energy induced by the different nucleon masses (Eqn.~(1c)) is of this order, 
and
is always taken into account, both in the $NN$ interaction 
and in the trinucleon
CSB, where it contributes a robust 
14 keV\cite{av18,brandenburg,sasakawa,argonne,nogga}. A
power-counting estimate of its size can be made by using $m_{\pi}$ to estimate
the average value of the nucleon momentum in the trinucleons, which produces a
value $\sim (m_{\pi}^2/M^2) \delta M_N \sim $ 25 keV (and within a
factor of two of the actual value). 
The size of this contribution illustrates the general
caveat about power counting:
even though this kinetic-energy
modification is an order smaller in the Lagrangian than the potential couplings
($n=3$ vs. $n=2$), they can make 
actual contributions in a ${\bf nucleus}$ that are not very different in size.
This contribution added to the soft-EM mechanisms discussed earlier
leaves about 73 keV to be explained by the various CSB nuclear potentials. 

The
other mechanism of order $n=3$ is the two-pion-exchange CSB arising from the
nucleon-mass difference in certain loops, which we discussed above and will
treat below. It depends not simply on the full
proton-neutron mass difference, but 
on $\delta M_N^{\rm qm}$ and $\delta M_N^{\rm EM}$ separately.
Since these two components of the nucleon-mass difference are 
presently unknown, we can only estimate the impact of
this potential on observable quantities.
Power counting suggests that this contribution is
one order smaller than the (three) previously discussed mechanisms. We can make
a numerical estimate of its contribution to the $^3$He $\minus$ $^3$H
binding-energy difference 
using naive dimensional analysis.
Again assuming that the typical momentum in the nucleus is $Q\sim m_\pi$
and that the loop integral gives the usual factor of $(4\pi)^{-2}$,
we expect in coordinate space
$V_{2 \pi}^{\rm CSB} \sim (m_{\pi}^4/64 \pi f_{\pi}^2 \Lambda^2)\, \delta M_N$,
which is more than an order of magnitude smaller than
the estimate for $\delta T$ of 25 keV. 
As we will see below, however, the loop integrals
in this case only give {\bf one} power of $(4 \pi)^{-1}$.
The CSB TPEP then likely supplies a larger contribution than expected
on the basis of power counting. 

Note that the CSB TPEP contribution is also a part of the traditional
``short-range''  nuclear CSB
mechanism, and its effect is therefore included in the 65(22) keV short-range
part of the trinucleon binding-energy difference. One can of course accommodate
all of these short-range mechanisms using models\cite{rho-w,csmodel}.
Alternatively one can hope that a sophisticated partial-wave analysis of
nucleon-nucleon scattering, such as that carried out by the Nijmegen
group\cite{rob_mart,nijmegen}, may be able to disentangle the CSB interactions
of one-pion range, two-pion range, and short range, each of which contains an
unknown parameter that must be fitted to the data. Incorporating these forces
into their procedure should be straightforward, although there may not be 
enough
sensitivity in the data to distinguish between the three mechanisms.
A full analysis of few-nucleon systems 
within effective field theory\cite{texas,improv,morenogga} 
will eventually include all these CSB mechanisms.

In summary, there are two mechanisms for CSB of order $n=2$ at the Lagrangian
level (CSB short-range forces in Eqn.~(2), and CSB OPEP in Eqn.~(3)) and one of
order $n=3$ (nuclear CSB kinetic energy in Eqn.~(1c)) that should make roughly
comparable contributions in a nucleus. The TPEP modified by 
$\delta M_N^{\rm qm}$ and $\delta M_N^{\rm EM}$,
together with the associated CSB seagull interactions 
(all given in Eqns.~(9)
below), should be somewhat smaller.

\section{CSB TPEP}

Isospin-conserving two-pion-exchange potentials (TPEPs) are an old problem 
with a
new twist. In static order (containing only terms that remain when the nucleon
mass, $M$, or the large-mass scale of QCD, $\Lambda$, becomes very large) the
diagrams of Fig.~(1) (properly symmetrized)
contribute to the TPEP. The vertices and 
propagators follow
from the leading-order Lagrangian for pions and nucleons,
$$
 {\cal L}^{(0)}  = \frac{1}{2}[\dot{\bpi}^{2}-(\vnabla \bpi)^{2}
          -\mpi^{2}\bpi^{2}] 
   + N^{\dagger}[i\partial_{0}-\frac{1}{4 \fpi^{2}} \btau \cdot
         (\bpi\times\dot{\bpi})]N +\frac{g_{A}}{2 \fpi} 
 N^{\dagger}\vsig \cdot\vnabla(\btau \cdot \bpi)N \, , \eqno (4)
$$
where the $\pi \pi N$ term is the Weinberg-Tomozawa (WT) interaction\cite{wt}
and the $\pi N$ term is the usual interaction that depends on the axial-vector
coupling constant, $g_A$, and the pion-decay constant, $\fpi$.  Terms with
additional pions or nucleons are neglected here, as they only contribute to the
nuclear force at higher orders. The WT term has a specific normalization ($-1 /
4 \fpi^2$) required by the underlying chiral symmetry.

\begin{figure}[tb]
\epsfig{file=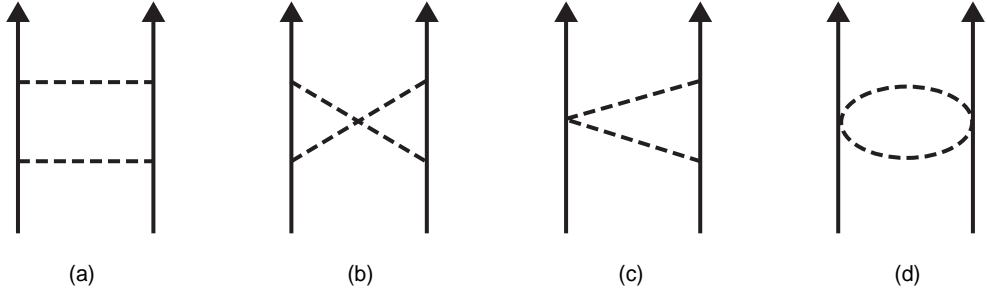,height=1.5in}
\caption{Two-pion-exchange graphs that contribute to isospin-conserving
nucleon-nucleon scattering.}
\end{figure}

The nucleon-mass difference corresponds to Lagrangian terms 
that arise from both
quark-mass differences, $\delta M_N^{\rm qm}$, and from EM interactions at the
quark level, $\delta M_N^{\rm EM}$. The largest term of the first type 
is\cite{iv}
$$
{\cal L}_{\rm qm}^{(1)} = - \delta M^{\rm qm}_N \; N^{\dagger}[\haf \tau_3 
- \frac{1}{4 \fpi^{2}} \pi_3 \btau \cdot \bpi ]N \, , \eqno (5)
$$
and the leading-order term of the EM type is\cite{iv}
$${\cal L}_{\rm EM}^{(-1)} = - \delta M^{\rm EM}_N \; N^{\dagger}[\haf \tau_3 
- \frac{1}{4 \fpi^{2}} (\bpi^2 \tau_3 - \pi_3 \btau \cdot \bpi)]N 
\, . \eqno (6)
$$
As with the WT term, the $\pi \pi N$ interactions 
in Eqns.~(5) and (6) are {\it required} by chiral symmetry
and have a fixed strength ($-1 /4 \fpi^2$) relative to the mass
terms. We again drop terms that involve more pion fields
and do not contribute to the nuclear potential in low orders.

The CSB TPEP can be computed in a straightforward way.
We simply consider all insertions of the mass and interaction
terms above into the diagrams of Fig.~(1), including external
lines. 
Alternatively,
we implement the $p$-$n$ mass-difference mechanism by including 
the simple $p$-$n$ mass
difference (i.e, the sum of the first terms in each Lagrangian in Eqns. (5) and
(6)) in the initial and final nuclear states, and then compensate for this
addition by a subtraction. That is, we write for a single nucleon
$$
\delta M_{\rm CSB} = \haf \delta M_N (\tau_z - \tau_z^0) \, , \eqno (7)
$$
where $\tau_z^0$ is the reference value of $\tau_z$ for that nucleon. The
expectation value of this contribution for all the nucleons will vanish, 
as does
the contribution from uncrossed box diagrams. The crossed-box diagram shown in
Fig.~(2a) and the triangle diagram in Fig~(2b) can also be easily computed. In
addition there are seagull terms involving two pions (from the second parts of
the two Lagrangians in Eqns.~(5) and (6)), 
which will generate triangle diagrams
(Fig.~(2c)), and in principle ``football'' diagrams (Fig.~(2d)). The latter
vanish for this problem because of the isospin symmetries of the WT
(antisymmetric) and CSB seagull (symmetric) vertices.

\begin{figure}[tb]
\epsfig{file=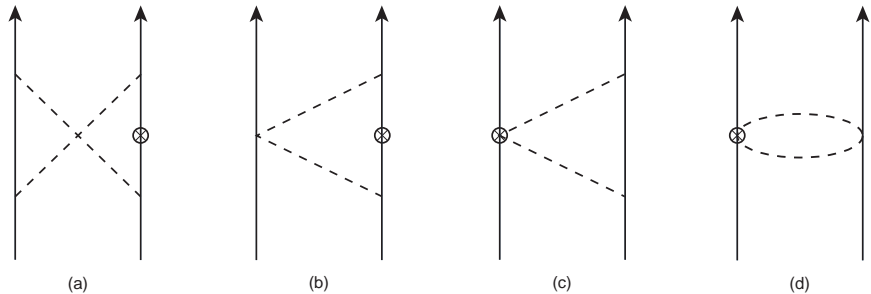,height=1.5in,bbllx=-100pt,bblly=0pt,bburx=415pt,
bbury=172pt}
\caption{Two-pion-exchange graphs that contribute to charge-symmetry breaking 
in
nucleon-nucleon scattering. Graph (d) vanishes because of the symmetry of
isospin operators. The $\otimes$ symbol indicates a CSB vertex, either a 
seagull
in (c) and (d), or a mass insertion in (a) and (b).}
\end{figure}

We find a result that is a pure class (III) potential.
Our results in momentum space are given below, with both pion momenta
$\vq_1 = \vk + \haf \vq$ and $\vq_2 = -\vk +\haf \vq$ pointing into the seagull
vertex. The loop momentum is $\vk$, and $\vq$ is the momentum transfer. 
Writing 
$$
V^{a,b,c}_{\rm CSB}(\vq\,) 
= \frac{1}{4\pi} 
  \left(\frac{g_A}{(2 f_{\pi})^2}\right)^2 \left(\tau_z(1)+\tau_z(2)\right) 
  \, v^{a,b,c}(\vq\,)  \eqno (8)
$$
and
$$
v^{a,b,c}(\vq\,) = 8\pi\, 
\int \frac{d^3k}{(2\pi)^3} \, u^{a,b,c}(\vq_1, \vq_2) \, , \eqno (9)
$$
graph~(2a) gives
$$
u^a(\vq_1, \vq_2) = - g_A^2 \, \delta M_N 
\, \left(\frac{1}{\vq_1^{\; 2} +m_{\pi}^2}
    + \frac{1}{\vq_2^{\; 2} + m_{\pi}^2}\right)
\frac{(\vq_1 \cdot \vq_2)^2
+\vsig(1) \cdot \vq_1 \times \vq_2 \; \vsig(2) \cdot \vq_1 \times \vq_2}
{(\vq_1^{\; 2} +m_{\pi}^2 ) (\vq_2^{\; 2} + m_{\pi}^2)} \, ,
\eqno (9a)$$
while graph (2b) generates
$$
u^b(\vq_1, \vq_2) = -\delta M_N 
\, \frac{\vq_1 \cdot \vq_2 }
{(\vq_1^{\; 2} +m_{\pi}^2 ) (\vq_2^{\; 2} + m_{\pi}^2)} \, ,\eqno (9b)
$$
and finally graph (2c) produces
$$
u^c(\vq_1, \vq_2) = \left(\delta M_N -\haf \delta M_N^{\rm qm}\right)
\, \frac{\vq_1 \cdot \vq_2 }
{(\vq_1^{\; 2} +m_{\pi}^2 ) (\vq_2^{\; 2} + m_{\pi}^2)} \,. \eqno (9c)
$$

The integrals in these expressions are divergent, requiring
regularization and renormalization by the
spin-independent nucleon-nucleon contact terms given in Eqn.~(2).
The loop integration over $\vk$ gives for the non-analytic terms
$$
\eqalignno{
v^a (\vq) = g_A^2  \, \delta M_N
&\left( 2\frac{q^2 + 2 m_{\pi}^2}{q} 
\arctan \left(\frac{q}{2 m_{\pi}} \right)  
+6 m_{\pi} 
-\frac{2m_{\pi}^3}{q^2+4 m_{\pi}^2} \right. 
& \cr 
&\left.- \frac{\vsig(1) \times \vq \cdot \vsig(2) \times \vq}{q}
\; \arctan \left(\frac{q}{2 m_{\pi}}\right) \right) \, , & (8a)\cr}
$$
and
$$
v^{b+c} (q) = - \delta M_N^{\rm qm}\,
\left( \frac{q^2 + 2 m_{\pi}^2}{2 q} 
\arctan \left(\frac{q}{2 m_{\pi}}\right) + m_{\pi} \right) 
\, . \eqno(8bc)
$$

The long-range part of the potential
in configuration space is independent of the regularization procedure.
It is simplest to derive it by introducing 
a cutoff (``form'') factor $F(\vq^{\; 2})$ for each pion line carrying
a momentum $\vq$ in or out of a vertex.
The three nucleon-nucleon potentials 
depicted in Figs.~(2a-c) are:
$$
V_{\rm CSB}(\vec{r}) = \frac{1}{(4 \pi)^2}
                         \left(\frac{g_A m_{\pi}^2}{(2 f_{\pi})^2}\right)^2 
                         \, (\tau_z(1)+\tau_z(2))\; v^{a,b,c}(m_\pi \vec{r})
\eqno(10) 
$$
with
$$
\eqalignno{
v^a(\vec{x}) = & 2g_A^2 \, \delta M_N \, \left(
(h(x) \vnabla^2 h(x) + x\, h^{\prime}(x) \vnabla^2 h(x) - 2(h^{\prime}(x))^2) 
\right. & \cr
& \left. + \vsig(1) \cdot \vsig(2) \; (h(x)\vnabla^2 h(x) - 
h(x) h^{\prime}(x)/x + (h^{\prime}(x))^2) \right. &\cr
&\left. -\vsig(1) \cdot \hat{r} \vsig(2) \cdot \hat{r} \;
(h(x) \vnabla^2 h(x) -3 h(x) h^{\prime}(x)/x +(h^{\prime}(x))^2)
\right)\, ,  &  (10a)\cr} 
$$
and
$$
v^{b+c}(\vec{x})= \delta M_N^{\rm qm} \, (h^{\prime}(x))^2 \, . \eqno(10bc) 
$$
In these potentials we have defined $\vx= m_{\pi} \vr$, 
all derivatives are with
respect to $x$, and $h(x)$ is
the regulated Yukawa function
$$
h (x) \equiv 4 \pi \int \frac{d^3l}{(2\pi)^3} \; 
\frac{F^2(\vl^{\; 2} m_{\pi}^2)}{(\vl^{\; 2} + 1)} 
e^{i {\vl} \cdot {\vx}} \rightarrow \frac{e^{-x}}{x}\, ,
\eqno (11)
$$
where the last form holds only for $F \equiv 1$. Note that $\vnabla^2 h(x)$
generates a purely short-range contribution (i.e., a $\delta$-function if
$F\equiv 1$) that is indistinguishable from other short-range contributions. It
is therefore permissible to ignore such terms and replace $\vnabla^2 h(x)$ by
$h(x)$.

Eqns. (10), (10a), and (10bc) form our main result: the model-independent,
long-range part of the CSB TPEP. This result can now be used,
for example, as added input to 
the Nijmegen phase-shift analysis\cite{rob_mart}.

The contribution from the crossed-box diagram, Eqn.~(9a),
is proportional to the total $p$-$n$ mass difference $\delta M_N$.
Our results 
agree with the functional form of Ref.\cite{csb1} in momentum space. A
potential in configuration space was not presented there.
The contribution from the the CSB seagull, Eqn.~(9c),
was first calculated in Ref.\cite{csb2}.
The triangle with mass insertion, Eqn.~(9b),
has not been previously calculated.
There are cancellations between the latter two terms, which
make them proportional to the quark-mass component $\delta M_N^{\rm qm}$ of the
nucleon-mass difference.
While the quantity $\delta M_N$ is a precisely known observable, 
the quantity $\delta M_N^{\rm qm}$ is not (yet).
Note that
our expressions (8a) and (8bc) contain only one power of $(4\pi)^{-1}$,
while usually a loop contributes two powers. 
Ignoring $\delta M_N^{\rm qm}$ and treating the dimensionless $h(x)$ and any 
of
its derivatives as order (1),
we get 
$V_{2 \pi}^{\rm CSB} \sim 
(m_{\pi}^4/8 f_{\pi}^2 \Lambda^2)\, \delta M_N$,
which is about half the estimate for $\delta T$ of 25 keV. 
Numerical estimates of the effect
of terms (a) and (c) have already been published\cite{csb1,csb2}. 
At present
only the effect of the sum of the one-pion, (the just-calculated) two-pion, and
short-range CSB potentials is constrained by experiment, while none of the
individual terms are uniquely specified.

In summary, we have calculated the additional two-pion-exchange CSB
nucleon-nucleon force that is the same order as (and complements) the
two-pion-exchange two-nucleon CSB force previously calculated in
Refs.\cite{csb1,csb2}. This force has been developed using ChPT, and should be
somewhat smaller than the other CSB interactions that we discussed. 
We have also
discussed the interplay between the CSB OPEP and short-range CSB force, both of
which are larger than the CSB TPEP. The observed CSB in the $^3$He
$\minus$ $^3$H system is consistent with modern calculations. No
model-independent resolution of the CSB nuclear potentials into components of
different range has yet been made, because each force of one-pion, two-pion, or
short range contains an (as yet) undetermined constant. It is hoped that by
completing the one-pion-range and two-pion-range parts of these potentials in
ChPT a phase-shift analysis of nucleon-nucleon scattering data that 
incorporates
this information may be able to differentiate the components of various ranges.
In addition to the two-pion-exchange nucleon-nucleon force derived here and in
Refs.\cite{csb1,csb2}, there will be two-pion-exchange three-nucleon CSB forces
of order $n=3$, which we have not treated herein.

\begin{center}
{\large {\bf Acknowledgments}}\\
\end{center}
We would like to thank Rob Timmermans and Bob Wiringa for several very helpful 
discussions about CSB.
UvK is grateful to the Department of Physics at the University
of Washington for its hospitality,
and to RIKEN, Brookhaven National Laboratory and the U.S.
Department of Energy [DE-AC02-98CH10886] for providing the facilities
essential for the completion of this work.
The work of JLF was performed under the auspices of the DOE, 
GLP was supported in part by the DOE, 
SAC was supported in part by the U.S. National Science Foundation,
and UvK was supported in part by the 
DOE Outstanding Junior Investigator Program
and the Alfred P. Sloan Foundation.


\begin{thebibliography}{999}

\bibitem{iv1} E.\ M.\ Henley, in {\it Isospin in Nuclear Physics}, D.\ H.\
         Wilkinson, ed.\ (North-Holland, Amsterdam, 1969), p.15; 
         E.\ M.\ Henley 
         and G.\ A.\ Miller, in {\it Mesons and Nuclei}, M.\ Rho and G.\ E.\ 
         Brown, eds. (North-Holland, Amsterdam, 1979), Vol.\ I, p.\ 405.
\bibitem{iv2} G.\ A.\ Miller, B.\ M.\ K.\ Nefkens, and I. \v{S}laus,
         {\it Phys.\ Rep.} {\bf 194}, 1 (1990); G.\ A.\ Miller and W.\ T.\ H.\ 
         van Oers, in {\it Symmetries and Fundamental Interactions in Nuclei}, 
         W.\ Haxton and E.\  M.\ Henley, eds. (World Scientific, Singapore, 
         1995).
\bibitem{iv3} S. A. Coon, {\tt nucl-th/9903033}, invited talk given at 
         XIII\underline{th} International Seminar on High Energy Physics
         Problems (ISHEPP 13), ``Relativistic Nuclear Physics and Quantum 
         Chromodynamics'', 
         Joint Institute for Nuclear Research, Dubna, Russia, September, 1996.
\bibitem{weinberg} S.\ Weinberg, {\it Physica} {\bf 96A}, 327 (1979); S.\ 
         Weinberg, {\it Nucl.\ Phys.} {\bf B363}, 3 (1991); {\it Phys.\ Lett.}
         {\bf B251}, 288 (1990); {\it Phys.\ Lett.} {\bf B295}, 114 (1992).
\bibitem{iv} U.\ van Kolck, Ph.\ D.\ Thesis, University of Texas, 1993; 
         U.\ van Kolck, {\it Few-Body Systems Suppl.} {\bf 9}, 444 (1995).
\bibitem{texas}  C.\ Ord\'o\~nez and U.\ van Kolck, {\it Phys.\ Lett.} 
        {\bf B291}, 459 (1992); C.\ Ord\'o\~nez, L.\ Ray, and U.\ van Kolck, 
        {\it Phys.\ Rev.\ Lett.} {\bf 72}, 1982 (1994); U. van Kolck, {\it 
        Phys. Rev. C} {\bf 49}, 2932 (1994); C.\ Ord\'o\~nez, L.\ Ray, and U.\ 
        van Kolck, {\it Phys.\ Rev.\ C} {\bf 53}, 2086 (1996).
\bibitem{2NTPE} J.\ L.\ Friar and S.\ A.\ Coon, 
                {\it Phys.\ Rev.} {\bf C49}, 1272 (1994);
                N.\ Kaiser, R.\ Brockmann, and W.\ Weise,
                {\it Nucl.\ Phys.}  {\bf A625}, 758 (1997); 
                J.-L.\ Ballot, M.\ R.\ Robilotta, and C.\ A.\ da Rocha,
                {\it Phys.\ Rev.} {\bf C57}, 1574 (1998);
                E.\ Epelbaoum, W.\ Gl\"ockle, and U.-G.\ Mei{\ss}ner,
                {\it Nucl.\ Phys.} {\bf A637}, 107 (1998); 
                J.\ L.\ Friar, 
                {\it Phys.\ Rev.\ C} {\bf 60}, 034002 (1999).
\bibitem{BvK} P.\ F.\ Bedaque and U.\ van Kolck, {\it Ann.\ Rev.\ Nucl.\ Part.
        \ Sci.} {\bf 52} (2002). This is a comprehensive current review of the
        application of ChPT to nuclear physics.
\bibitem{ndpc}  A.\ Manohar and H.\ Georgi, {\it Nucl.\ Phys.} {\bf B234}, 189 
        (1984); H.\ Georgi, {\it Phys.\ Lett.} {\bf B298}, 187 (1993).
\bibitem{dpc} J.\ L.\ Friar, {\it Few-Body Systems} {\bf 22}, 161 (1997).
\bibitem{nijmegen} M.\ C.\ M.\ Rentmeester, R.\ G.\ E. Timmermans,
        J.\ L.\ Friar, and J.\ J.\ de Swart, {\it Phys.\ Rev.\ Lett.} {\bf 82},
        4992 (1999).
\bibitem{rob_mart} R.\ G.\ E. Timmermans (Private Communication). The Nijmegen
         group is presently analyzing the effect of chiral two-pion-exchange 
         nuclear forces on $np$ scattering. Their $pp$ results are given in 
         Ref.~\cite{nijmegen}.
\bibitem{av18} R.\ B.\ Wiringa, V.\ G.\ J.\ Stoks, and R.\ Schiavilla, 
        {\it Phys. Rev. C} {\bf 51}, 38 (1995).
\bibitem{pig} U.\ van Kolck, M.\ C.\ M.\ Rentmeester, J.\ L.\ Friar, 
        T.\ Goldman, and J.\ J.\ de Swart, {\it Phys.\ Rev.\ Lett.} {\bf 80}, 
        4386 (1998).
\bibitem{1loop} U.\ van Kolck, J.\ L.\ Friar, and T.\ Goldman, 
         {\it Phys.\ Lett.} {\bf B371}, 169 (1996).
\bibitem{FvK} J.\ L.\ Friar and U.\ van Kolck, {\it Phys.\ Rev.\ C} {\bf 60}, 
        34006 (1999).
\bibitem{csb1} S.\ A.\ Coon and J.\ A.\ Niskanen, 
        {\it Phys.\ Rev.\ C} {\bf 53}, 1154 (1996). 
\bibitem{csb2} J.\ A.\ Niskanen, {\it Phys.\ Rev.\ C} {\bf 65},
        037001 (2002). 
\bibitem{cs-rev} R.\ Machleidt, {\it Adv.\ Nucl.\ Phys.} {\bf 19}, 189 
        (1990); G.\ A.\ Miller, {\it Nucl.\ Phys.} {\bf A578}, 345 (1990); 
        {\it Chinese J.\ Phys.} {\bf 32}, 1075 (1994). These are fairly recent 
        reviews of charge symmetry. See also Refs.\cite{iv1,iv2,iv3}.
\bibitem{Allena} A.\ K.\ Opper and E.\ Korkmaz (spokespersons), 
TRIUMF E-704 Proposal.
\bibitem{EdAndy} A.\ D.\ Bacher and E.\ J.\ Stephenson (spokespersons), 
IUCF CE-82 Proposal.
\bibitem{vKMN} U.\ van Kolck, J.\ A.\ Niskanen, and G.\ A.\ Miller,
{\it Phys.\ Lett.} {\bf B493}, 65 (2000).
\bibitem{olderlit} D.\ O.\ Riska and Y.\ H.\ Chu,
                   {\it Nucl.\ Phys.} {\bf A235}, 499 (1974);
                   J.\ V.\ Noble, in {\it The Interaction Between Medium Energy
                   Nucleons in Nuclei} (AIP Conf. Proc. 97), H.-O.\ Meyer, ed.
                   (AIP, New York, 1983), p. 83;
                   P.\ G.\ Blunden and M.\ J.\ Iqbal,
                   {\it Phys.\ Lett.} {\bf B385}, 25 (1996).
\bibitem{pi-eta} S.\ A.\ Coon and M.\ D.\ Scadron, {\it Phys.\ Rev. C} 
        {\bf 26}, 562 (1982).
\bibitem{coulomb} J.\ L.\ Friar, B.\ F.\ Gibson, and G.\ L.\ Payne, {\it 
        Phys.\ Rev.\ C} {\bf 35}, 1502 (1987).
\bibitem{brandenburg} R.\ A.\ Brandenburg, S.\ A.\ Coon, and P.\ U.\ Sauer, 
        {\it Nucl.\ Phys.} {\bf A294}, 305 (1978).
\bibitem{sasakawa} Y.\ Wu, S.\ Ishikawa, and T.\ Sasakawa, 
        {\it Phys.\ Rev.\ Lett.} {\bf 64}, 1875 (1990); {\bf 66 (E)}, 
        242 (1991).
\bibitem{argonne} S.\ C.\ Pieper, V.\ R.\ Pandharipande, R.\ B.\ Wiringa, and 
      J.\ Carlson, {\it Phys.\ Rev.\ C} {\bf 64}, 014001 (2001) [see Table XI];
      S.\ C.\ Pieper and R.\ B.\ Wiringa, {\it Ann.\ Rev.\ Nucl.\ Part.\ Sci.}
        {\bf 51}, 53 (2001).
\bibitem{nogga} 
A.\ Nogga, A.\ Kievsky, H.\ Kamada, W.\ Gl\"ockle, L.\ E.\  Marcucci, 
S.\ Rosati, and M.\ Viviani, nucl-th/0202037.
\bibitem{csmodel} R.\ Machleidt and H.\ M\"uther, {\it Phys.\ Rev. C} {\bf 63},
        034005 (2001); G.\ Q.\ Li and R.\ Machleidt, {\it Phys.\ Rev. C} 
        {\bf 58}, 1393 (1998).
\bibitem{rho-w} S.\ A.\ Coon and R.\ C.\ Barrett, {\it Phys.\ Rev. C} 
        {\bf 36}, 2189 (1987).
\bibitem{a1-f1} S.\ A.\ Coon, B.\ H.\ J.\ McKellar, and V.\ G.\ J.\ Stoks,
        {\it Phys.\ Lett.} {\bf B385}, 25 (1996).
\bibitem{improv} E.\ Epelbaum, W. Gl\"ockle, and U.-G.\ Mei{\ss}ner,
                {\it Nucl. Phys.} {\bf A671}, 295 (2000);
                D.\ R.\ Entem and R. Machleidt, 
                {\it Phys. Lett.} {\bf B524}, 93 (2002).
\bibitem{morenogga}
E.\ Epelbaum, A.\ Nogga, W.\ Gl\"ockle, H.\ Kamada, U.-G.\ Mei{\ss}ner, 
and H.\ Wita\l a, {\it Phys.\ Rev. C} {\bf 66}, 064001 (2002). 
\bibitem{wt} S.\ Weinberg, {\it Phys.\ Rev.\ Lett.} {\bf 17}, 616 (1966);
        Y.\ Tomozawa, {\it Nuovo Cimento A} {\bf 46}, 707 (1966).

\end{thebibliography}
\end{document}